\documentclass[11pt]{article}

\usepackage{fullpage}

\usepackage{amsmath,amssymb,theorem}

\title{On the Quantum Black-Box Complexity of Majority}

\date{}

\author{
Thomas P. Hayes\footnote{Department of Mathematics, University of
Chicago, 5734 S.\ University Avenue, Chicago, IL 60637.
Email: {\tt hayest@math.uchicago.edu}.}
 \and Samuel Kutin\footnote{Department of Computer Science,
University of Chicago, 1100 E.\ 58th Street, Chicago, IL 60637.
Email: {\tt kutin@cs.uchicago.edu}.}
\and Dieter van Melkebeek\footnote{Computer Sciences Department,
University of Wisconsin, 1210 W.\ Dayton Street, Madison, WI 53706.
Email: {\tt dieter@cs.wisc.edu}.
Supported in part by NSF Career award CCR-0133693 and NSF grants
EIA-0130400 and EIA-0205236.}}

\newcommand{\XOR}{{\rm XOR}}

\newcommand{\PARITY}{{\rm PARITY}}
\newcommand{\PAR}{{\mathcal{PARITY}}}
\newcommand{\MAJORITY}{{\rm MAJORITY}}

\newcommand{\A}{{\cal A}}
\newcommand{\G}{{\cal G}}
\newcommand{\T}{{\cal T}}
\newcommand{\CS}{{\cal S}}

\newcommand{\zo}{ \{0,1\} }

\newcommand{\into}{\rightarrow}
\newcommand{\xor}{\oplus}
\newcommand{\qb}[1]{\left|{#1}\right>}

\newcommand{\floor}[1]{\left\lfloor{#1}\right\rfloor}
\newcommand{\ceil}[1]{\left\lceil{#1}\right\rceil}
\newcommand{\Dxor}{D^{\rm XOR}}
\newcommand{\Dpar}{D^{\PAR}}

\newcommand{\subroutine}{COMBINE}

\def\proof{\noindent{\it Proof.} \ }
\newcommand{\qed}[1][]{\hfill #1 $\square$ \bigskip}

{\theorembodyfont{\rmfamily}
       \newtheorem{definition}{Definition}  }

\theoremstyle{plain}
    \newtheorem{theorem}[definition]{Theorem}
    \newtheorem{corollary}[definition]{Corollary}
    \newtheorem{lemma}[definition]{Lemma}

{\theorembodyfont{\rmfamily}
	}

\begin{document}

\maketitle

\begin{abstract}
We describe a quantum black-box network computing the majority
of $N$ bits with zero-sided error $\epsilon$ using only 
$\frac{2}{3}N + O(\sqrt{N \log (\epsilon^{-1} \log N)})$ queries:
the algorithm returns the correct answer with probability at least
$1 - \epsilon$, and ``I don't know'' otherwise.
Our algorithm is given as a randomized ``{\XOR} decision tree''
for which the number of queries on any input is strongly concentrated
around a value of at most $\frac{2}{3}N$.
We provide a nearly matching lower bound of $\frac{2}{3}N - O(\sqrt N)$
on the expected number of queries on a worst-case input
in the randomized {\XOR} decision tree model with zero-sided error
$o(1)$. Any classical randomized decision tree computing the majority
on $N$ bits with zero-sided error $\frac{1}{2}$ has cost $N$.
\end{abstract}

\section{Introduction}

How do you tell how a committee of three people will vote on an
issue?  The obvious approach is to ask each individual what vote
he or she is planning to cast.  If the first two committee members
agree, you can skip the third one, but, if they disagree, you need
to talk to all three members.

Suppose, however, that you can perform quantum tranformations on
the committee members.  This allows you to ask, with one quantum
question, whether the first two members agree or disagree.
If they agree, you can
disregard the third member and ask one of the first two for her
vote.  If the first two disagree, you know their votes will cancel,
so it suffices to ask the third member for his vote.  Either way,
you will learn the answer in only two queries.

In this paper, we discuss generalizations of this procedure to
arbitrarily many voters.  We allow our algorithms to ask whether
two voters agree at the cost of one query.  We consider both
deterministic and randomized algorithms, allowing different kinds
of error.  Our algorithms can be simulated very efficiently on
quantum machines, yielding new upper bounds for the quantum
complexity of the {\MAJORITY} function.

\subsection{Overview}
 
Suppose we wish to compute the value $f(X)$ of a
function $f$ on $\zo^N$ where the input $X$ is given to us
as a {\em black-box} $X\colon\{0$, $\dots$, $N-1\}\into\zo$.
The cost of the computation will be the number of queries we make to
the oracle $X$. In the classical case, this model of computation is
known as a {\em decision tree}, and has been well-studied.

More recently, a {\em quantum mechanical} version of the model has
been considered, which is inherently probabilistic. 
Several complexity measures are investigated:
the number of queries needed to compute $f$ exactly, with zero-sided
error $\epsilon$, or with bounded error $\epsilon$.
Beals et al.~\cite{BBCMdW} show that for any function $f$ these
measures are all polynomially related to the classical decision tree
complexity. Beals et al.\ also look more closely at some specific
functions $f$. In particular, they consider the majority function,
whose decision tree complexity equals $N$. They prove that in the
quantum model the exact and zero-sided error cost functions are 
between $N/2$ and $N$ (for any $\epsilon < 1$); a result of 
Paturi's \cite{paturi} implies that the bounded error cost
function is $\Omega(N)$ (for any constant $\epsilon < \frac{1}{2}$).

In this paper, we investigate these cost measures for {\MAJORITY} more
closely. We provide improved upper bounds, as well as matching
lower bounds in related models.

Our first result is a quantum black-box network which exactly
computes {\MAJORITY} using $N+1-w(N)$ queries, where $w(N)$ equals 
the number of
ones in the binary expansion of $N$. So, for $N$ of the form
$2^n-1$, we can save $\floor{\log N}$ queries.

Our algorithm exploits the fact, due to Cleve et al.~\cite{CEMM},
that the {\XOR} of two input bits can be determined in a single quantum
query. In fact, our algorithm can be viewed as an {\XOR} decision tree,
i.e., a classical
decision tree with the additional power of computing the {\XOR} of two
input bits
at the cost of a single query.  The complexity of {\MAJORITY} in this model
has been studied before~\cite{SW91, ARS93, ARS97}, independently
of the connection with quantum computation. A tight bound of
$N+1-w(N)$ was known~\cite{SW91, ARS93}.
We give a simpler proof for the lower bound which generalizes to the 
case where computing the parity of arbitrarily many input bits is permitted 
in one query. The lower bound shows that our procedure cannot 
be improved without at least introducing a new quantum trick.

Our main result is a quantum black-box network that computes
{\MAJORITY} with zero-sided error $\epsilon$ using only 
$\frac{2}{3}N + O(\sqrt{N \log(\epsilon^{-1} \log N)})$ queries. For 
any positive $\epsilon$ we construct such a network. The 
algorithm can be viewed as a randomized variant of an {\XOR} decision 
tree given by Alonso et al.~\cite{ARS97}. We construct an exact 
randomized {XOR} decision tree with an expected number of queries of at 
most $\frac{2}{3}N + 2 \log N$ on any input. We argue that the number of
queries is sufficiently concentrated to yield our main result.

Alonso et al.~\cite{ARS97} show that the {\em average} cost of their 
algorithm over all $N$-bit inputs is $\frac{2}{3}N - \Omega(\sqrt N)$.  
They also show that the average-case complexity of {\MAJORITY} in 
the {\XOR} decision tree model is at least $\frac{2}{3}N - O(\sqrt N)$.  
We instead are interested in the cost of randomized {\XOR} decision trees 
on {\em worst-case} inputs. A standard argument shows that
the Alonso et al.\ lower bound also holds for the expected number of
queries on a worst-case input. We also prove that classical randomized
decision trees need $N$ queries to compute {\MAJORITY} with zero-sided
error $\frac{1}{2}$.

In the general bounded-error setting, Van Dam~\cite{vD98} has shown how
to compute any function $f$ using $\frac{1}{2}N + \sqrt{N \log \epsilon^{-1}}$ 
quantum queries. 
We point out that Van Dam's technique does not provide a zero-sided 
error network for {\MAJORITY} of cost less than $N$.
We prove that any
classical randomized decision tree for {\MAJORITY} has to have cost
$N$ to achieve bounded error of at most $\frac{1}{4}$.

\subsection{Organization}

Section~\ref{sec:prelim} provides some preliminaries, including
background on the {\XOR} decision tree model, the
quantum black-box model, and their relationship.
Section \ref{sec:exact}
describes and analyzes our quantum network for computing
{\MAJORITY} exactly using $N+1-w(N)$ queries.
In Section~\ref{sec:zero-error}, we discuss our randomized 
{\XOR} decision tree for {\MAJORITY} that has small zero-sided error and
cost about $\frac{2}{3}N$, and we relate this to the zero-error quantum 
query complexity.
In Section~\ref{sec:lower:exact}, we show that the
exact algorithm of Section~\ref{sec:exact} is optimal in a generalized
version of the {\XOR} decision tree model.
In Section~\ref{sec:lower:zero}, we discuss lower bounds for the
cost of randomized {\XOR} decision trees and classical randomized
decision trees for computing {\MAJORITY}.
Finally, in Section~\ref{sec:questions}, we give a table summarizing 
the known results and propose several questions for further research.

\section{Preliminaries} \label{sec:prelim}

We first introduce some general notation. Then we discuss 
{\XOR} decision trees, quantum black-box networks, and their
relationship.

Let $X = X_0 X_1 \ldots X_{N-1}$ be a Boolean string of length $N$. We
will often think of $X$ as a function $X: \{0,1,\ldots,N-1\} \rightarrow
\zo$. We define $\MAJORITY(X)$ to be 0 if $X$ contains more zeros than 
ones, and 1 otherwise. This is a weak definition, which we will use to
establish our lower bounds. Our algorithms will always yield a stronger
result in that they will answer ``tie'' when the number of zeros
and ones are equal. The {\it discrepancy} of $X$ is the size of the
majority, i.e., the absolute value of the difference in the number of
zeros and ones. $\XOR$ denotes the exclusive OR of two bits, and
$\PARITY(X)$ denotes $\sum X_i \bmod 2$. 

For a positive integer $N$, the {\em Hamming weight} of $N$, denoted
$w(N)$, is the number of ones in the standard binary representation
for $N$. We will use the following properties.
\begin{lemma}\label{hamming-lemma}
For any integer $N > 0$, $\sum_{k=1}^{\infty} \floor{N/2^k} = N -
w(N)$.
\end{lemma}
\proof
Let $\ell = \floor{\log N}$, and write $N = \sum_{j=0}^\ell b_j 2^j$,
where $b_j \in\zo$.
We then have:
$$
\sum_{k=1}^\infty \floor{N/2^k} =
\sum_{k=1}^\infty \sum_{j=k}^\ell b_j 2^{j-k} =
\sum_{j=1}^\ell b_j \sum_{k=1}^j 2^{j-k} =
\sum_{j=1}^\ell b_j (2^j - 1) =
\sum_{j=0}^\ell b_j 2^j - \sum_{j=0}^\ell b_j
$$
which is simply $N - w(N)$.\qed
\begin{corollary} \label{hamming-corollary}
For any integer $N > 0$, $N!$ is exactly divisible by $2^{N - w(N)}$.
\end{corollary}
\proof
For any positive integer $k$, there are exactly $\floor{N/2^k}$ multiples of 
$2^k$ contributing to $N!$.  So the exponent of the largest power of 2
dividing $N!$ is given by
$\sum_{k=1}^\infty \floor{N/2^k}$, which is equal to
$N - w(N)$ by Lemma~\ref{hamming-lemma}. \qed

\subsection{{\XOR} decision trees}
\label{sec:prelim-dtrees}

An {\em XOR decision tree} is an algorithm for a given input length $N$
which adaptively queries the input $X$ and outputs a value.
A query may be either:
\begin{itemize}
\item $X_i$, where $0 \le i \le N-1$, or
\item $X_i \oplus X_j$, where $0 \le i,j \le N-1$ and $\oplus$
denotes $\XOR$.
\end{itemize}
The cost on a given input $X$ is the number of queries made.
The cost of an {\XOR} decision tree is the maximum cost over all inputs
of length $N$. An {\XOR} decision tree can be viewed as a binary tree.
The depth of this tree equals the cost of the {\XOR} decision tree.
We refer to Section \ref{sec:tree:gen} for a further generalization
of {\XOR} decision trees.

We define a {\em randomized XOR decision tree} $\T$ as an {\XOR} decision
tree in which we can toss a coin with arbitrary bias at any point in 
time, and proceed based on the outcome of the coin toss. Equivalently,
we can view $\T$ as a probability distribution over (deterministic) 
{\XOR} decision trees. The number of queries on a given input $X$ is a 
random variable. We define the {\em cost} on input $X$ as the maximum 
of this random variable, and the cost of $\T$ as the maximum cost over 
all inputs $X$. 

The following definitions applies to a randomized decision tree $\T$ on
$N$-bit inputs, and more generally to any probabilistic process $\T$ that takes
a Boolean string of length $N$ as input and outputs a value.
Let $f$ be a function on $\zo^N$. If on any input $X$, $\T$ outputs
$f(X)$ with probability at least $1 - \epsilon$, we say that $\T$ computes
$f$ with error $\epsilon$. If $\T$ outputs $f(X)$ with probability at least 
$1 - \epsilon$ and says ``I don't know'' otherwise (i.e., $\T$ never produces
an incorrect output) we say that $\T$ computes $f$ with {\em zero-sided error}
$\epsilon$. In the case where $\epsilon = 0$, we say that $\T$ {\em exactly} computes 
$f$. 

A randomized decision tree that exactly computes $f$ at cost $C$
can trivially 
be transformed into a deterministic tree computing $f$ at the same cost. 
It can often also be transformed into a randomized {\XOR} decision tree 
for $f$ with zero-sided error $\epsilon$ and cost $C' < C$, e.g.,
if on any input the number of queries is strongly concentrated around
a value less than $C'$. More precisely, suppose that on any input $X$, 
with probability at least $1-\epsilon$, $\T$ makes no more than $C'$ queries. 
Then we can run $\T$ but as soon as we attempt to make more than $C'$
queries, stop the process and output ``I don't know.'' The modified
randomized decision tree has zero-sided error at most $\epsilon$ and
cost at most $C'$.

\subsection{Quantum black-box networks}
\label{sec:prelim-quantum}

A quantum computer performs a sequence of unitary transformations
$U_1, U_2, \ldots, U_T$ on a complex Hilbert space, called the {\em
  state space}. The state space has a canonical orthonormal basis 
which is indexed by the configurations $s$ of some classical 
computer $M$. The basis state corresponding to $s$ is denoted by
$\qb{s}$.

The initial state $\phi_0$ is a basis state. At any point in time $t$,
$1 \leq t \leq T$, the state $\phi_t$ is obtained by applying 
$U_t$ to $\phi_{t-1}$, and can be written as
\[
\phi_t = \sum_s \alpha_{s,t} \qb{s}
\]
where $\sum_s |\alpha_{s,t}|^2 = 1$.

At time $T$, we {\em measure} the state $\phi_T$.  This is a
probabilistic process that produces a basis state, where the probability of
obtaining state $\qb{s}$ for any $s$ equals $|\alpha_{s,T}|^2$. 
The output of the algorithm is the observed state $\qb{s}$ or some
part of it. 

We define the {\em quantum black-box model} following Deutsch and 
Jozsa~\cite{DJ92}. In a quantum black-box network $\A$ for input length $N$, 
the initial state $\phi_0$ is independent of the input 
$X = X_0 X_1 \ldots X_{N-1}$. 
We allow arbitrary unitary transformations independent of $X$. 
In addition, we allow $\A$ to make {\em quantum queries}. This is the
transformation $U$ taking the basis state $\qb{i,b,z}$ to $\qb{i, b\xor
  X_i, z}$, where:
\begin{itemize}
\item $i$ is a binary string of length $\log N$ denoting an index into
  the input $X$,
\item $b$ is the contents of the location where the result of the
  oracle query will be placed,
\item $z$ is a placeholder for the remainder of the state description,
\end{itemize}
and comma denotes concatenation.

We define the {\em cost} of $\A$ to be the number of times the query
transformation $U$ is performed; all other transformations are free.

The error notions introduced in Section \ref{sec:prelim-dtrees} for
arbitrary probabilistic processes also apply to quantum black-box
networks.

\subsection{From {\XOR} decision trees to quantum black-box networks}
\label{sec:prelim-XtoQ}

Bernstein and Vazirani have shown~\cite{BV97}
that a quantum computer can efficiently simulate classical
deterministic and probabilistic computations. It is also known that we 
can efficiently compose quantum algorithms. In terms of quantum
black-box networks these results imply that a classical randomized
decision tree $\T$ that uses quantum black-box networks as subroutines 
can be efficiently simulated by a single quantum black-box network. 
The cost of the simulation will be the sum of the cost of $T$ and the
costs of the subroutines. Similarly, the error of the simulation will 
be bounded by the sum of the error of $\T$ and the errors of the 
subroutines. The simulation will have zero-sided error if all of
the components do.

We will describe our quantum black-box networks for
{\MAJORITY} as classical randomized decision trees that use the
following exact quantum black-box network developed by Cleve 
et al.~\cite{CEMM} for computing the XOR of two input bits.
\begin{lemma}[Cleve et al.~\cite{CEMM}] \label{thm:XOR}
There exists a quantum black-box network of unit cost that on input 
two bits $X_0$ and $X_1$ exactly computes their XOR.
\end{lemma}

The above argument shows that an {\XOR} decision tree for a function
$f$ can be transformed into a quantum black-box network for $f$ of the
same cost. The transformation works in the exact setting, as well as 
for zero-sided or arbitrary error $\epsilon$.

\section{Computing {\MAJORITY} Exactly}
\label{sec:exact}

In the introduction, we discussed how to use an {\XOR} query to
determine the {\MAJORITY} of three input bits.  In this section,
we generalize this idea to an input of arbitrary length.
We first describe a general approach for constructing
{\XOR} decision trees or exact randomized {\XOR} decision trees for 
\MAJORITY. We call it the ``homogeneous block approach.'' We
use this approach to develop the ``oblivious-pairing'' algorithm, an 
{\XOR} decision tree that computes 
$\MAJORITY$ exactly on $N$-bit inputs using at most $N+1-w(N)$ queries. In 
Section \ref{sec:lower:exact} we will show that this is optimal.

The oblivious-pairing algorithm was first introduced and analyzed by 
Saks and Werman~\cite{SW91}. It forms a first step towards the
zero-sided error randomized {\XOR} decision tree for {\MAJORITY} which we 
will develop in Section \ref{sec:zero-error}.

\subsection{The homogeneous block approach}

{\XOR} queries allow us to compare bits of the input $X$. If the bits
differ in value, we can discard them since the two of them together will 
not affect the majority value. If the bits have the same value, we can 
combine them into a homogeneous block of size 2, i.e., a subset of 2
input bits which we know have the same value but we do not know what
that value is. More generally, we can apply the following operation
``\subroutine'' to two disjoint nonempty homogeneous blocks $R$ and $S$.
Suppose that $|R| \geq |S|$. We compare a bit from $R$ with a bit from 
$S$. If the bits differ, we discard block $S$ completely together with
$|S|$ bits from block $R$.  Otherwise, we combine blocks $R$ and $S$ 
into a single homogeneous block of size $|R|+|S|$. 

In the homogeneous block approach, we keep track of a collection of
disjoint nonempty homogeneous blocks with the property that the 
majority of the bits in the union of the blocks equals $\MAJORITY(X)$.
We start out with the partition of the input into blocks of size 1,
i.e., individual bits. Then we use some criterion to decide to which
two blocks we apply the operation {\subroutine}. We keep doing so until we
end up in a configuration consisting of an empty collection or one
in which one of the blocks is larger than the union of all other blocks.
In the former case, we have a tie. In the latter, the largest block
determines the majority, and querying any of its bits gives us the value
of the majority. One of these situations will eventually be reached since 
the number of blocks goes down by 1 or 2 in each step.

Building a homogeneous block of size $k$ requires only $k-1$ comparisons
between the bits in the block. In general, the number of comparisons
performed upon reaching a configuration consisting of $\ell$ homogeneous 
blocks equals $N - \ell - c$, where $c$ denotes the number of times two blocks
cancelled each other out completely. It follows that, compared to the 
trivial procedure of querying every input bit, the homogeneous block 
approach saves one query for every block in the final configuration 
except the dominating block, and one for every cancellation of equal-sized
blocks.

\subsection{The oblivious-pairing algorithm}

In the oblivious-pairing algorithm, we first build homogeneous blocks of 
size 2 by pairing up the initial blocks of size 1, leaving the last block 
of size 1 untouched when $N$ is odd. Then we build blocks of size 4 out 
of the blocks of size 2, possibly leaving the last block of size 2 
untouched, etc. In general, during the $k$th phase of the algorithm, we
will pairwise \subroutine\ the homogeneous blocks of size $2^{k-1}$ to either
cancel or form homogeneous blocks of size $2^k$. There will be at most
one block of size $2^{k-1}$ left after the end of the $k$th phase. 

There can be at most $\floor{\log N}$ phases. Afterwards, either there are 
no blocks left, in which case we have a ``tie,'' or else all remaining 
blocks have sizes that are different powers of 2. The largest block then
dominates all the others combined and dictates the majority.

We provide pseudo-code for the oblivious-pairing algorithm in Figure
\ref{fig:OP}. We keep track of the collection of disjoint nonempty 
homogeneous blocks as a list $\CS \doteq (S_j)_{j=1}^{\ell}$ of subsets 
of $\{0,1,\ldots,N-1\}$ of nonincreasing size. We will always compare two 
consecutive blocks in the list, say $S_i$ and $S_{i+1}$, a procedure captured
by the subroutine \subroutine. We also use the following
notation: If $X$ is homogeneous on a subset $S$ of $\{0,1,\ldots,N-1\}$, 
we write $X_S$ for the value of any bit $X_i$, $i \in S$.

\begin{figure}
\centerline{
\fbox{
\begin{minipage}{100in}
\addtolength{\baselineskip}{.3\baselineskip}
\begin{tabbing}
    \underline{input:} $X \doteq (X_i)_{i=0}^{N-1} \in \{0,1\}^N$ \\
    \underline{output:} $\MAJORITY(X)$ \\
    \underline{notation:} \= $\ell \doteq |\CS|$ \\
		          \> $S_j \doteq$ $j$th element of $\CS$, 
					$1 \leq j \leq \ell$ \\
			  \> $X_{S_j} \doteq X_i$ for any $i \in S_j$, 
					$1 \leq j \leq \ell$ \\
    \underline{subroutine:} \= \subroutine($\CS$, $i$, $X$) \+\\
	{\bf if} \= $\XOR(X_{S_i}, X_{S_{i+1}}) = 0$ \+ \\
	     {\bf then} replace $S_i$, $S_{i+1}$ in
                        $\CS$ by $S_i \cup S_{i+1}$ \\
             {\bf else} remove $S_i$, $S_{i+1}$ from $\CS$\-\-\\
    \underline{algorithm:} \\
    algo\=\+\kill
        $\CS \leftarrow (\{i\})_{i=0}^{N-1}$ \\
        {\bf for} \= $k = 1,2,\ldots,\floor{\log N}$ \+ \\
          {\bf while} $I \doteq \{ j \, | \, 1 \leq j < \ell \mbox{ and }
                          |S_j|=|S_{j+1}|=2^{k-1} \} \not= \emptyset$ \\
          {\bf for} \= \+\kill
                $i \leftarrow \min I$ \\
		COMBINE($\CS,i,X$) \-\- \\
        {\bf if} \= $\ell = 0$ \+\\
             {\bf then} return ``tie'' \\
             {\bf else} return $X_{S_1}$
\end{tabbing}
\end{minipage}
}
}
\caption{The oblivious-pairing algorithm}
\label{fig:OP}
\end{figure}

For any positive integer $k$, the blocks of size $2^{k-1}$ are pairwise 
disjoint. We pair them up during the $k$th phase of the algorithm.
It follows that the number of \subroutine\ operations during the $k$th phase
is bounded from above by $\floor{N/2^k}$. Each application of \subroutine\ 
involves one {\XOR}. Therefore, Lemma \ref{hamming-lemma} gives us an upper 
bound of $N-w(N)$ on the total number of {\XOR}s. There can be at 
most one more query, for a total of $N+1-w(N)$. This total is reached, 
e.g., for homogeneous inputs (all zeros or all ones). There are no 
cancellations on homogeneous inputs, and $w(N)$ is the smallest number of 
power-of-2 blocks that add up to $N$. We conclude:

\begin{theorem}[Saks-Werman \cite{SW91}]
\label{exactMAJORITY-thm} \label{thm:exact}
The oblivious-pairing algorithm for {\MAJORITY} on $N$-bit inputs has
{\XOR} decision tree cost $N+1-w(N)$.
\end{theorem}

\begin{corollary}
We can compute {\MAJORITY} exactly on $N$-bit inputs using at most
$N + 1 - w(N)$ quantum black-box queries.
\end{corollary}

%
%

\section{Computing {\MAJORITY} with Zero-Sided Error}
\label{sec:zero-error}

In Section~\ref{sec:exact}, we considered the oblivious-pairing {\XOR}
decision tree. We showed that it has a cost of $N - w(N) + 1$.
We now consider exact randomized {\XOR} decision trees for {\MAJORITY}.
Our main result is the randomized greedy-pairing algorithm, for which 
the number of queries on any input is highly concentrated around a value 
of about $\frac{2}{3}N$ on a worst-case input.
Using the techniques discussed in
Sections \ref{sec:prelim-dtrees} and \ref{sec:prelim-XtoQ}, this gives
us a randomized {\XOR} decision tree and a quantum black-box network with
small zero-sided error of cost about $\frac{2}{3}N$.
In Section~\ref{sec:lower:zero}, we will give a
nearly matching lower bound on the expected number of queries on a
worst-case input for randomized {\XOR} decision trees with small zero-sided 
error. 

In Section~\ref{sec:expected-random}, we discuss a simple randomized 
version of the oblivious-pairing algorithm. We carefully analyze the
number of queries it makes, as we will need that result later on.
In Section~\ref{sec:expected-ARS}, we describe a
deterministic algorithm of Alonso, Reingold, and Schott~\cite{ARS97},
the greedy-pairing algorithm, for which the average number of queries
over all $N$-bit inputs is roughly $\frac{2}{3}N$.  
In Section~\ref{sec:expected-random-ARS}, we analyze a randomized 
version of the greedy-pairing algorithm. We prove that the number
of queries it makes is with high probability not much larger than
$\frac{2}{3}N$.

\subsection{The randomized oblivious-pairing algorithm}
\label{sec:expected-random}

The oblivious-pairing algorithm is efficient when we can get pairs of 
blocks to cancel.  Recall that the number of {\XOR}s made in any
homogeneous block algorithm for \MAJORITY\ equals $N-\ell-c$, where
$\ell$ denotes the number of blocks at the end, and $c$ the number
of cancellations of equal-sized blocks that occurred. In the
oblivious-pairing algorithm, $\ell$ can be at most $\log N$, so not
much savings can be expected from that term. The number of
cancellations can be much larger. On the input $010101\ldots$,
all $N/2$ pairs of individual bits cancel, and we can declare a tie with
only $N/2$ queries. However, even if we know the input is perfectly
balanced, there is no guarantee that any cancellations occur until the
very end.

One natural approach is to randomly permute the input bits before
we begin the algorithm: Choose some permutation $\pi$ of $\{0,1,\ldots,
N-1\}$ uniformly at random, let $X'_i = X_{\pi(i)}$, and run the 
oblivious-pairing algorithm on the input $X'$.  The distribution of
the number of queries on a given input now only depends on the number 
of ones and the number of zeros it contains. 

Consider the randomized oblivious-pairing algorithm running on
a perfectly balanced input of length $N$. We perform $N/2$ queries 
comparing individual bits; we expect roughly half of those to cancel, 
and half to yield homogeneous blocks of size 2. We next pair up the 
$N/4$ blocks of size 2, which takes $N/8$ queries. Again, we expect 
roughly half of those queries to cancel, and half to yield blocks of 
size 4.  The overall number of queries should then be about
$$\frac{N}2 + \frac{N}8 + \frac{N}{32} + \dots = \frac{2}{3}N.$$
We prove below that the number of queries the oblivious-pairing 
algorithm makes on a balanced input is indeed highly concentrated
around $\frac{2}{3}N$.

However, consider a homogeneous input. Permuting the input bits has no 
effect; the input remains homogeneous, blocks will never cancel, and 
the randomized oblivious-pairing algorithm still takes $N - w(N) + 1$ 
queries.  We will need to do something else to reduce the computation 
cost on such inputs.  We return to this question in 
Section~\ref{sec:expected-ARS}.

Before doing so, we prove the following theorem about the number
of comparisons the oblivious-pairing algorithm makes on input
$X$. We will use the theorem in our analysis of our main result in 
Section~\ref{sec:expected-random-ARS}.

\begin{theorem}\label{thm:OP}
There exists a constant $d$ such that the following holds.
Let $C_{OP}(X)$ denote the number of comparisons the oblivious-pairing 
algorithm makes on input $X$.  Let $N > 0$, and let $A+B=N$,
$A, B \ge 0$.
Let $X$ be chosen uniformly at random
from all strings of $A$ ones and $B$ zeros.
Then for any $r \geq 1$,
\[ 
\Pr_X \left[ \, C_{OP}(X) \geq N - \frac{2}{3}\min(A,B) +  d \sqrt{rN} 
      \, \right]  \leq 2^{-r} \log N. 
\]
\end{theorem}

The proof of Theorem \ref{thm:OP} uses the following tail law. 

\begin{lemma}\label{lemma:OP}
There exists a constant $d'$ such that the following holds.
Let $c(X)$ denote the number of cancellations during the first phase
of the oblivious-pairing algorithm on input $X$.  Let $N > 0$, and let $A+B=N$,
$A, B \ge 0$.
Let $X$ be chosen uniformly at random from all strings of $A$
ones and $B$ zeros.
Then for every $r \geq 1$,
\[   \Pr_X \left[ \, |c(X) - AB/N| \geq d' \sqrt{rN} \, \right] \leq 2^{-r}. \]
\end{lemma}

The combinatorial problem underlying Lemma \ref{lemma:OP} is a special case of
``Levene's matching problem'' \cite{david-barton}, and has been 
well studied. We suspect that the tail law given in Lemma \ref{lemma:OP}
is known but have not been able to find a reference. We include a proof 
in the Appendix.

\medskip

\noindent
{\it Proof of Theorem \ref{thm:OP}}.\ \ 
The proof goes by induction on $N$. We first do the induction step. 

Assume without loss of generality that $A \geq B$. 
Look at the sequence of homogeneous blocks of size 2 after the first 
phase of oblivious-pairing on input $X$. Let $X'$ denote the input obtained 
by replacing each block in this sequence by a single bit of the same value. 
We have that $C_{OP}(X) = \floor{\frac{N}{2}} + C_{OP}(X')$. 

Let $A'$ denote the number of ones in $X'$, $B'$ the number of zeros, and
$N' = A' + B'$. Note that $N' = \floor{\frac{N}{2}} - c(X)$,
$B' = \floor{\frac{B-c(X)}{2}}$, and $A' \geq B'$. 

Conditioned on $A'$ and $B'$, the distribution of $X'$ is uniform. Therefore,
by our induction hypothesis, we have that with probability at least
$1 - 2^{-r} \log N'$
\begin{eqnarray*}
C_{OP}(X') & \leq & N' + \frac{2}{3} B' + d \sqrt{rN'} \\
           &  =   & \floor{\frac{N}{2}} - c(X) - \frac{2}{3} 
			\floor{\frac{B-c(X)}{2}}
			+ d \sqrt{rN'} \\
           & \leq & \frac{N}{2} - \frac{B}{3} - \frac{2}{3}c(X)
			+ d \sqrt{rN'} + \frac{1}{3}.
\end{eqnarray*}
By Lemma \ref{lemma:OP}, with probability at least $1-2^{-r}$,
\[ c(X) \geq AB/N - d' \sqrt{rN} \geq \frac{B}{2} - d' \sqrt{rN}. \]
Taking everything together, and using that fact that $N' \leq N/2$, 
we have that with probability at least
$1 - 2^{-r}\log N' - 2^{-r} \geq 1 - 2^{-r}\log N$,
\begin{eqnarray*}
 C_{OP}(X) & \leq & N - \frac{2}{3}B + (\frac{2d'}{3} + \frac{d}{\sqrt{2}})
			\sqrt{rN} + \frac{1}{3} \\
	   & \leq & N - \frac{2}{3}B + d \sqrt{rN},
\end{eqnarray*}
provided $d$ is large enough that 
$\frac{2d'+1}{3} \leq (1-\frac{1}{\sqrt{2}})d$. This proves the induction
step.

By picking $d$ larger as needed, we can take care of the base cases.
\qed

Theorem \ref{thm:OP} can be strengthened to show that the random variable
$C_{OP}(X)$ is strongly concentrated around a value slightly smaller
than $N - \frac{2}{3}\min(A,B)$. We omit the precise expression for the 
concentration point, as it is rather cumbersome and not needed for the
sequel. A proof similar to the above (but simpler and not relying on Lemma 
\ref{lemma:OP}) shows that the expected value of $C_{OP}(X)$ in 
Theorem \ref{thm:OP} is bounded above by $N - \frac{2}{3} \min(A,B)$.

\subsection{The greedy-pairing algorithm}
\label{sec:expected-ARS}

As we mentioned in Section~\ref{sec:expected-random}, the
oblivious-pairing algorithm requires $N - w(N) + 1$ queries
on the all ones input, whether or not we randomize. In contrast,
the trivial algorithm for {\MAJORITY}, which simply queries
bits until the observed discrepancy is larger than the number
of bits remaining, takes $\floor{N/2}+1$ queries on the
all ones input. Therefore, we should be able to improve the
oblivious-pairing algorithm.

The oblivous-pairing algorithm always {\subroutine}s two smallest
blocks of equal size.
A first idea is that we may decide to always {\subroutine} two
{\em largest} blocks of equal size instead, and stop as soon as
the largest block (if any) is larger than the union of the other
blocks.
This leads to an improvement on some inputs, e.g., on 
homogeneous inputs of length $N = 2^k - 1$: we will build up a 
block of size $2^{k-1}$ using $\floor{N/2}$ {\XOR}s and query one 
bit in that block, for a total cost of $\floor{N/2}+1$. However, 
on homogeneous inputs of length $N = 2^k + 1$, we still make
$N-1$ queries: we construct a block $S_1$ of size $2^{k-1}$, and
then perform another $2^{k-1} - 1$ queries to form another large 
block, even though one additional query combining $S_1$ with another 
bit would guarantee a majority. 

In order to do better, we should allow {\subroutine} operations
on blocks of unequal size. As cancellations of blocks of equal
size are beneficial, we will still prefer to {\subroutine} such 
blocks, but we should only do so if we reasonably expect the answer 
to be useful. Alonso, Reingold, and Schott~\cite{ARS97} introduce
a homogeneous block algorithm for {\MAJORITY} which does just this: 
They {\subroutine} two blocks only if they are sure they will need 
to know the answer. We call this the ``greedy-pairing'' algorithm.

More precisely, the greedy-pairing algorithm works as follows. 
Suppose that in some step we find a pair $S_i$, $S_{i+1}$ of large 
blocks of equal size. Instead of automatically combining these two blocks, 
however, we
now ask a question: Are we sure this is necessary? In other words, 
if we assumed all blocks up to $i$ all agreed, would that still not 
be enough to determine a majority?  If the answer is yes, we 
{\subroutine} the two blocks. If the answer is no, then we try to
build up the largest block by running {\subroutine} on $S_1$ and $S_2$.

When we compare two blocks of the same size, we are trying to gain
by cancelling and reducing $\ell$ by 2 in a single step.  When we
compare two blocks of different sizes, we are trying to gain by greedily 
constructing a large enough block to guarantee a majority.

Since the only {\subroutine} operations between blocks of unequal size
involve $S_1$, all blocks except possibly $S_1$ will have sizes that
are powers of 2. Say $|S_j| = 2^{s_j}$, $2 \leq j \leq \ell \doteq |\CS|$, 
where the $s_j$'s are integers. The size of $S_1$ can be written as
$|S_1| = (2m+1) 2^{s_1}$ for some integers $m$ and $s_1$. Note that
$s_1 \geq s_2 \geq \ldots \geq s_\ell$. We will think of $S_1$ as 
being composed of several power-of-2 blocks. The smallest such subblock
has size $2^{s_1}$. 

The precise criterion we use to determine which blocks $S_i$ and $S_{i+1}$
to compare is given in the pseudo-code of Figure \ref{fig:GP}. Note that
the smallest $j$ such that $s_j = s_{j+1}$ exists during each execution
of the while loop. If there were no such $j$, the block $S_1$ would 
dominate all the other blocks combined and we would have exited the loop.

\begin{figure}
\centerline{
\fbox{
\begin{minipage}{100in}
\addtolength{\baselineskip}{.3\baselineskip}
\begin{tabbing}
    \underline{input:} $X \doteq (X_i)_{i=0}^{N-1} \in \{0,1\}^N$ \\
    \underline{output:} $\MAJORITY(X)$ \\
    \underline{notation:} \= $\ell \doteq |\CS|$ \\
		          \> $S_j \doteq$ $j$th element of $\CS$, 
					$1 \leq j \leq \ell$ \\
			  \> $X_{S_j} \doteq X_i$ for any $i \in S_j$, 
					$1 \leq j \leq \ell$ \\
                          \> $s_1 \doteq$ largest integer $t$ such that
 				$2^t$ divides $|S_1|$ \\
                          \> $s_j \doteq |S_j|$, $2 \leq j \leq \ell$ \\
    \underline{subroutine:} \= \subroutine($\CS$, $i$, $X$) \+\\
	 {\bf if} \= $\XOR(X_{S_i}, X_{S_{i+1}}) = 0$ \+ \\
		{\bf then} replace $S_i$, $S_{i+1}$ in
			$\CS$ by $S_i \cup S_{i+1}$ \\
		{\bf else} {\bf if} \= $|S_i| > |S_{i+1}|$ \+ \\
			{\bf then} \= remove $|S_{i+1}|$ elements from $S_i$\\
				   \> remove $S_{i+1}$ from $\CS$ \\
			{\bf else} remove $S_i$, $S_{i+1}$ from $\CS$ \-\-\-\\
    \underline{algorithm:} \\
    algo\=\+\kill
        $\CS \leftarrow (\{i\})_{i=0}^{N-1}$ \\
         {\bf while} \= $\ell > 0$ and $|S_1| \leq \sum_{j=2}^\ell |S_j|$ \+ \\
           $i$ $\leftarrow $ smallest integer $j$ such that $s_j$ = $s_{j+1}$\\
           {\bf if} \= $\sum_{j=1}^i |S_j| > \sum_{j=i+1}^\ell |S_j|$ \\
	            \> {\bf then} $i$ $\leftarrow$ $1$ \\
		COMBINE($\CS,i,X$) \- \\
        {\bf if} \= $\ell = 0$ \+\\
             {\bf then} return ``tie'' \\
             {\bf else} return $X_{S_1}$
\end{tabbing}
\end{minipage}
}
}
\caption{The greedy-pairing algorithm}
\label{fig:GP}
\end{figure}

The key to the good performance of the greedy-pairing algorithm is the
following observation. 
Let $M$ denote the index of the $(\floor{N/2} + 1)$st input bit
agreeing with the majority. If $X$ is balanced, let $M \doteq N$. Let
$Y$ denote the substring consisting of the first $M$ bits of $X$, and
$Z$ the remainder of $X$.
Then the greedy-pairing algorithm never performs any comparisons
involving bits of $Z$.
This is because $Y$ forces the majority in all of $X$, and the
greedy-pairing algorithm only involves a new bit $b$ in a comparison if 
the bits before $b$ cannot force the majority of $X$.

This is the way the greedy-pairing algorithm saves queries compared to the 
oblivious-pairing algorithm: by not making the comparisons the 
oblivious-pairing algorithm makes involving bits of $Z$. On $Y$, the 
greedy-pairing algorithm makes some of the comparisons the oblivious-pairing
algorithm makes, but possibly also makes some others. We need to show that 
there aren't too many other queries, or at least that we can account for 
most of them by queries the oblivious-pairing algorithm makes on $Y$ but the
greedy-pairing algorithm does not. We will prove next that there are
at most $O(\log^2 N)$ queries that we cannot account for in that way.

\begin{theorem}\label{thm:GPvsOP}
Let $C_{GP}(X)$ denote the number of comparisons the greedy-pairing
algorithm makes on input $X$, and let $C_{OP}$ be defined as in Theorem
\ref{thm:OP}.
There exists a constant $d$ such that on any binary input $X$ of length
$N$,
\[ C_{GP}(X) \leq C_{OP}(Y) + d \log^2 N, \]
where $Y$ denotes the first $M$ bits of $X$ and $M$ the position of the
$(\floor{N/2} + 1)$st bit in $X$ agreeing with the majority.
When $X$ is balanced, $M \doteq N$ and $Y \doteq X$.
\end{theorem}

In fact, a refinement of the argument below shows that
\[ C_{OP}(Y) \leq C_{GP}(X) \leq C_{OP}(Y) + \max(2 \floor{\log N} - 3, 0),\]
which is tight. However, the relationship as stated in Theorem
\ref{thm:GPvsOP} is strong enough for our purposes.

In order to prove Theorem \ref{thm:GPvsOP}, we need the following properties
of the greedy-pairing algorithm. They deal with the technical concept of an 
``unusual comparison,'' which is a comparison between $S_1$ and $S_2$ with 
$s_1 \not= s_2$. These are precisely the comparisons between blocks of
different sizes, provided we view a comparison with $S_1$ as one with
the last subblock of $S_1$ of size $2^{s_1}$.

\begin{lemma}\label{lemma:GP}
Consider running the greedy-pairing algorithm on an input $X$ and
call a comparison unusual if it is between $S_1$ and $S_2$, and
$s_1 \not= s_2$. Let $s$ be an integer. Let $T$ be the first point in time 
there is an unusual comparison with $s_2 \leq s$. (If there is no such
comparison, we let $T$ denote the end of the 
algorithm.) Then the following hold:
\begin{enumerate}
\item All comparisons the greedy-pairing algorithm makes before $T$ with 
     $|S_{i+1}| \leq 2^s$ are also made by the oblivious-pairing
     algorithm on input $X$.
\item After $T$, the greedy-pairing algorithm makes no comparisons 
     with $|S_{i+1}| \geq 2^s$ and $i>1$, and none with $|S_{i+1}| > 2^s$ 
     and $i=1$.
\item\label{it} Let $B_j$ denote the $j$th block $S_2$ of size $2^s$ which 
     the greedy-pairing algorithm
     compares with $S_1$ at and after $T$. Then the sequence $B_1, B_2,
     \ldots, B_r$ are successive blocks of size $2^s$ produced by the
     oblivious-pairing algorithm on input $X$.
\item The outcome of each of the greedy-pairing comparisons referred to in 
     \ref{it} is "unequal" for $S_2 = B_j$, $1 \leq j < r$.
\end{enumerate}
\end{lemma}

\noindent
{\it Proof of Lemma \ref{lemma:GP}.}\ \
We prove claim 1 by contradiction.  Suppose that, at some time before $T$,
the greedy algorithm makes a comparison with $|S_{i+1}| = 2^u$, where
$u \le s$, which is not made by the oblivious-pairing algorithm.
Consider the first such time $U$.  Since $U < T$, the comparison at
time $U$ is not unusual.  Since no unusual comparisons with
$|S_{i+1}| \le 2^u$ have occurred, we must have $|S_i| = |S_{i+1}|$.
And, by our choice of $U$, both $S_i$ and $S_{i+1}$ are also formed
by the oblivious-pairing algorithm.

By our choice of $U$, any earlier blocks of size $2^u$ must have been
compared as they are in the oblivious-pairing algorithm.  In
particular, there must be an even number of them.  So, blocks $S_i$ and
$S_{i+1}$ must be the $j$th and $(j+1)$st blocks of size $2^u$ formed by
the oblivious-pairing algorithm for some odd $j$.  Hence, this comparison
is also made by the oblivious-pairing algorithm, contradicting our
choice of $U$.

We now consider claim 2.  Clearly, at time $T$, only
$|S_1|$ can have size larger than $2^s$, and, if $s_2 > 0$,
$|S_i| < |S_2|$ for $i > 2$.  (Proof:  if $|S_2| = |S_3|$,
then $S_3$ must have been formed at some time $U < T$;
since the algorithm did not do an unusual comparison at time $U$, it
would have chosen to compare $S_2$ and $S_3$ at time $U+1$.)

At time $T$, let $j$ be the smallest index such that $|S_j| = |S_{j+1}|$.
Then we must have $\sum_{k=1}^j |S_k| > \frac12 \sum_{k=1}^\ell |S_k|.$
Since all blocks up to $S_j$ have different sizes, and
$|S_2| \le 2^s$, we conclude that, at time $T$,
\begin{equation}
\label{eqn-lemma:gp}
|S_1| + 2^{s+1} - 1 > \frac12 \sum_{k=1}^\ell |S_k|.
\end{equation}
Once inequality (\ref{eqn-lemma:gp}) holds, it remains true for
the remainder of the algorithm.  (No
comparison can increase the right-hand side.  The left side is
decreased only by an ``unequal'' comparison between $S_1$ and $S_2$,
in which case both sides decrease by $|S_2|$.)

So, suppose that, at some later time, there is a block of size $2^s$,
and, if a comparison were done between some $S_i$ and $S_{i+1}$, it
would form a second such block.  By (\ref{eqn-lemma:gp}), the
greedy-pairing algorithm would choose to do a comparison between
$S_1$ and $S_2$ instead.  Hence, from time $T$ onward, there can
be at most one block $S_i$ of size $2^s$ for $i>1$, which proves 
claim 2.

To prove claim 3, we let $U$ be the first time (if any) that there is an
unusual comparison with $s_2 < s$.  (If $s_2 < s$ at time $T$, then
$U = T$ and $r = 0$.)  By claim 2, all blocks $B_j$ are formed before
time $U$.  So, by claim 1 applied to $s-1$, the comparisons which form
those blocks are all performed by the oblivious-pairing algorithm.

Finally, by the above reasoning, at the time that $S_1$ is compared
to $B_j$, there is no other block of size $2^s$.  If the comparison
were ``equal,'' then the next comparison would be unusual as well,
with $|S_2| < 2^s$, and no additional blocks $B_j$ would form.
We conclude that, for each $j < r$, the comparison between $S_1$
and $B_j$ is ``unequal.''  
\qed

Using Lemma \ref{lemma:GP}, we can prove Theorem \ref{thm:GPvsOP} as
follows.

\medskip

\noindent
{\it Proof of Theorem \ref{thm:GPvsOP}.}\ \
Fix a nonnegative integer $s$ and look at the comparisons the
greedy-pairing algorithm makes on input $X$ with $|S_{i+1}|=2^s$. Let $T$
be as defined in Lemma \ref{lemma:GP}. 

By claim 1 of Lemma \ref{lemma:GP}, 
all such comparisons before $T$ are also 
made by the oblivious-pairing algorithm on input $X$ at some point in time. 
As the greedy-pairing algorithm only involves bits of $Y$ in comparisons, 
these comparisons are actually made by the oblivious-pairing algorithm on 
input $Y$.

By claims 2 and 3 of Lemma \ref{lemma:GP}, there are $r$ more comparisons 
the greedy-pairing algorithm makes at and after time $T$ with
$|S_{i+1}|=2^s$. With these comparisons, we can associate the comparisons 
the oblivious-pairing algorithm makes involving the blocks $B_1, B_2, 
\ldots, B_r$ and their superblocks. By claim 3,
$B_1, B_2, \ldots,
B_{r-1}$ are subsequent blocks the oblivious-pairing algorithm produces
during phase $s$. By claim 4, all of them have the same value.
The oblivious-pairing
algorithm will spend at least $r-1-\ceil{\log(r-1)}$ comparisons on combining
the blocks $B_1, \ldots, B_{r-1}$.
So, the greedy-pairing algorithm makes at most $r - (r-1-\ceil{\log(r-1)}) 
\leq 2 + \log N$ queries with $|S_{i+1}|=2^s$ which we cannot 
account for by queries the oblivious-pairing algorithm makes on $Y$.
Adding this surplus over all values of $s$ we get that
\[ C_{GP}(X) \leq C_{OP}(Y) +  (2 + \log N) \log N, \]
which establishes the upper bound.
\qed

We point out that Alonso et al.~\cite{ARS97} showed that the average 
case complexity of the greedy-pairing algorithm is optimal up to an 
$O(\log N)$ term. In particular, they established the following upper bound.
\begin{theorem}[Alonso et al.~\cite{ARS97}]
\label{thm:ARS}
The average number of comparisons made by the greedy-pairing algorithm
over all $N$-bit inputs equals
$$
\frac{2N}3 - \sqrt{\frac{8N}{9\pi}} + O(\log N).
$$
\end{theorem}
The term $\sqrt{8N/9\pi}$ comes from the average discrepancy over all 
$N$-bit inputs, which is $\sqrt{2N/\pi} + O(1)$.

However, the analysis by Alonso et al.\ is not sufficient for our purposes.
We need an algorithm which performs well on the worst-case input.  It is
with this goal in mind that we now study a randomized version of the 
greedy-pairing algorithm. 

\subsection{The randomized greedy-pairing algorithm}
\label{sec:expected-random-ARS}

In Section~\ref{sec:expected-random}, we randomized the
oblivious-pairing algorithm by first applying a random permutation
$\pi$ to the input bits.  We can use the same technique to randomize 
the greedy-pairing algorithm. This is the algorithm which leads to
our main result.

The following analysis is essential.

\begin{theorem}\label{thm:GP}
There exists a constant $d$ such that the following holds.
Let $C_{GP}(X)$ denote the number of comparisons the greedy-pairing
algorithm makes on input $X$. Let $N > 0$, and let $A+B=N$,
$A, B \ge 0$.
Let $X$ be chosen uniformly at random
from all strings of $A$ ones and $B$ zeros.
Then for every $r \geq 1$,
\[ \Pr_X \left[ \, C_{GP}(X) \geq 
          \frac{1}{2}N + \frac{1}{3}\min(A,B) + d \sqrt{rN} \, \right]
        \leq 2^{-r} \log N. \]
\end{theorem}

Theorem \ref{thm:GP} shows that the worst-case inputs for the randomized
greedy-pairing algorithm are the balanced ones. We conclude:

\begin{corollary}\label{cor:GP}
There exists a constant $d$ such that for any positive $\epsilon$ and any
binary string $X$ of length $N$, with probability at least
$1 - \epsilon$ the randomized greedy-pairing algorithm makes no more than 
$\frac{2}{3}N + d \sqrt{N \log(\epsilon^{-1} \log N)}$ comparisons
on input $X$.
\end{corollary}

As with Theorem~\ref{thm:OP}, we will make use of a
concentration result in the proof of Theorem~\ref{thm:GP}.
Here as there, we believe this result is already known, 
but have not found a reference.  A proof is included in
the Appendix.

\begin{lemma}\label{lem:Mconcentration}
There exists a constant $d$ such that the following holds.
Let $N = A+B > 0$, where $A \geq B \geq 0$.
Let $X$ be chosen uniformly at random from all strings of $A$ ones and
$B$ zeros.
Let $M$ denote the index of the $(\floor{N/2} + 1)$st one of $X$.
If $A=B$, let $M \doteq N$.
Then for every $r \geq 1$,
\[
\Pr_X \left[ \left| M - \frac{N^2}{2A} \right| > d\sqrt{rN} \right] \leq 2^{-r}\
.
\]
\end{lemma}

\noindent
{\it Proof of Theorem \ref{thm:GP}.} \ 

Let $Y$ be the string consisting of the first $M$ bits of $X$,
where, as before, $M$ is the position of the
$(\floor{N/2} + 1)$st bit in $X$ agreeing with the majority.
When $X$ is balanced, $M \doteq N$ and $Y \doteq X$.
By Theorem~\ref{thm:GPvsOP}, 
\[
C_{GP}(X) \leq C_{OP}(Y) + O(\log^2 N).
\]

If $X$ is exactly balanced, then $Y=X$ is uniformly distributed
among strings having $N/2$ ones and $N/2$ zeros.  In this case,
we have reduced to Theorem~\ref{thm:OP}.  

Suppose $X$ is not exactly balanced.  Without loss of generality,
let $A > B$.  
$Y$ has exactly $(\floor{N/2} + 1)$ ones and $M-(\floor{N/2} + 1)$
zeros, the $M$'th bit being a one.  Let $Y'$ be the string of length
$M-1$ obtained by dropping the last one from $Y$.

Conditioned on $M$ being fixed, 
$Y'$ is uniformly distributed among strings having
$\floor{N/2}$ ones and $M-1-\floor{N/2}$ zeros.
Hence Theorem~\ref{thm:OP} applied to $Y'$ yields
\[
C_{OP}(Y')
\leq M - \frac{2}{3}(M-\floor{N/2}) + d\sqrt{rM}
\leq \frac{M + N}{3} + d \sqrt{rN}
\]
with probability at least $2^{-r}\log N$.

Since $Y$ differs from $Y'$ only in the rightmost bit, oblivious
pairing does all the same comparisons on $Y$ as on $Y'$, plus at
most one additional comparison per phase.  Hence
\[
C_{OP}(Y) \leq C_{OP}(Y') + \log N.
\]

Putting this together, 
\[
C_{GP}(X) \leq \frac{M + N}{3} + d \sqrt{rN} + O(\log^2 N)
\]
with probability at least $1 - 2^{-r}\log N$.
Since $M \leq N$, this is already enough to
establish Corollary~\ref{cor:GP}.  

By Lemma~\ref{lem:Mconcentration},
$M \leq \frac{N^2}{2A} + d'\sqrt{rN}$ with probability at least $1-2^{-r}$.
Hence, with probability at least $1 - 2^{-r}(1 + \log N)$,
\begin{eqnarray*}
C_{GP}(X) &\leq& \frac{N^2 + 2AN}{6A} + (d + d')\sqrt{rN} + O(\log^2 N) \\
&=& \frac{3AN + BN}{6A} + d'' \sqrt{rN} \\
&\leq& \frac{3AN+2AB}{6A} + d'' \sqrt{rN} \\
&=& \frac{N}{2} + \frac{B}{3} + d'' \sqrt{rN}.
\end{eqnarray*}
\qed

Theorem \ref{thm:GP} can be strengthened to show that the random variable
$C_{GP}(X)$ is strongly concentrated around a value slightly smaller
than $\frac{N}{2} - \frac{\min(A,B)}{3}$. 
We omit the precise expression for the
concentration point, as it is rather cumbersome and not needed for the
sequel. 

A simplified version of the proof of Theorem \ref{thm:GP} shows that
the expected number of comparisons the randomized greedy pairing
algorithm makes on an $N$-bit input with $A$ ones and $B$ zeros is
bounded above by
\[ \frac{1}{2}N + \frac{1}{3}\min(A,B) + 2 \log N 
    = \frac{2}{3} N - \frac{1}{6} D + 2 \log N, \]
where $D$ denotes the discrepancy of the input.  This gives us a bound
of the form $\frac{2}{3}N - O(\sqrt N)$ on the average-case cost
of the (randomized) greedy-pairing algorithm. However, the constant hidden in the 
$O(\sqrt N)$ term is not as good as that achieved by
Alonso et al.~\cite{ARS97} in Theorem
\ref{thm:ARS}. 

Using the techniques from Section \ref{sec:prelim-XtoQ}, Corollary
\ref{cor:GP} yields our main result.
\begin{theorem}[Main Result]\label{thm:main}
There exists a constant $d$ such that, for any 
positive integer $N$ and any $\epsilon > 0$, there exists a quantum
black-box network of cost 
$$\frac{2}{3}N + d \sqrt{N \log(\epsilon^{-1} \log N)}$$
that computes the majority of $N$ bits with zero-sided error $\epsilon$.
\end{theorem}

\section{Lower Bounds for Computing {\MAJORITY} Exactly}
\label{sec:lower:exact}

Beals et al.~\cite{BBCMdW} establish a lower bound of $\frac{N}{2}$
quantum queries for computing {\MAJORITY} exactly. In this section, we
show that any {\XOR} decision tree computing {\MAJORITY} must use
at least $N+1-w(N)$.  Hence, the oblivious-pairing algorithm of
Section~\ref{sec:exact} is optimal.

We first define a more general model of computation, a
decision tree relative to a set of functions.  We then show
that, relative to the collection of all parity functions, the 
oblivious-pairing algorithm is the best possible.

Recall that
the classical decision tree complexity of {\MAJORITY} equals $N$.

\subsection{Relative decision trees complexity}\label{sec:tree:gen}

A decision tree relative to a class of functions $\G$ is one which
is permitted to
apply any function from $\G$ to a subset of the input bits (taken in
any order) at unit cost.
\begin{definition}[$\G$-decision tree]
Let $\G = \{g_1, g_2, \ldots\}$ be a collection of 
functions where $g_k$ is a function on $M_k$ bits. A {\em
  $\G$-decision tree} is a deterministic algorithm for a given input
length $N$ which can query its
input bits $X_0$, $\dots$, $X_{N-1}$, and which can also perform
queries of the form $g_k(X_{\sigma(0)}, \ldots, X_{\sigma(M_k-1)})$, 
where $\sigma$ is a one-to-one function from $\{0, \dots, M_k-1\}$ to 
$\{0, \ldots, N-1\}$.  The cost of a $\G$-decision tree is the maximum over
all $N$-bit inputs of the total number of queries performed on
that input, including individual input bits as well as functions $g_k$.
\end{definition}

\begin{definition}[$\G$-decision tree complexity]
Let $f$ be a function on $\{0,1\}^N$.  The {\em $\G$-decision tree 
complexity of $f$}, denoted $D^\G(f)$, is the minimum cost of a 
$\G$-decision tree computing $f$.  When $\G = \{g\}$, we write this 
simply as $D^g(f)$.
\end{definition}

We will consider two instances, namely $\G = \{ \XOR \}$ and
$\G = \PAR$, where $\PAR$ denotes the collections of all
{\PARITY} functions (on any number of bits). 

We trivially have that $\Dpar(f) \le \Dxor(f)$ for any function $f$.
The discussion in Section \ref{sec:prelim-XtoQ} shows that there
exists a quantum black-box network that computes $f$ exactly with
cost at most $\Dxor(f)$.

The following lemma establishes a limit on how much we can expect
$\PAR$ to help simplify the computation of a function $f$.
It is an extension of a result of Rivest and Vuillemin~\cite{RV76}
for standard decision trees.

\begin{lemma}
\label{rv-lemma}
Let $f$ be a Boolean function on $\zo^N$.  If $\Dpar(f) \le d$,
then $2^{N-d}$ divides $|f^{-1}(1)|$.
\end{lemma}

\proof
Each leaf of the decision tree corresponds to a set of inputs:
those inputs for which the computation terminates at that leaf.
These sets partition $\zo^N$; in particular, the accepting
leaves partition $f^{-1}(1)$.  So it suffices to prove that the
size of the set corresponding to any leaf is divisible by $2^{N-d}$.

View $\zo^N$ as a vector space of dimension $N$ over ${\rm GF}(2)$ (with
coordinate-wise addition).  Each parity query or input bit query is
of the form: ``Is the input in a subspace of codimension 1?''
(A subspace has {\em codimension\/} $c$ if it has dimension
$N - c$.)  If every
response is ``yes,'' then the set corresponding to the leaf is
also a subspace; since at most $d$ questions were asked, this
space is of codimension at most $d$.
If some response is ``no,'' then the set is an affine subspace.
This is either empty, or nonempty of codimension at most $d$.
In every case, the size of the set is a multiple of $2^{N-d}$.
\qed

\subsection{Lower bound for {\MAJORITY}}

As we have noted, the oblivious-pairing algorithm in Section~\ref{sec:exact}
is an {\XOR} decision tree.
Theorem~\ref{exactMAJORITY-thm} therefore
implies that $\Dxor(\MAJORITY) \leq N + 1 - w(N)$.  We now show
that equality actually holds.  Hence, the oblivious-pairing
algorithm is optimal.

\begin{theorem}
$\Dpar(\MAJORITY) = \Dxor(\MAJORITY) = N + 1 - w(N)$.
\end{theorem}

\proof
As noted above, we already know that $$\Dxor(\MAJORITY) \leq N + 1 - w(N)$$
by Theorem~\ref{exactMAJORITY-thm}.  Since $\Dpar(\MAJORITY) \leq
\Dxor(\MAJORITY)$, it suffices to show that $$\Dpar(\MAJORITY) \geq N
+ 1 - w(N).$$ We will use Lemma~\ref{rv-lemma} to do so; the first step
is to compute what power of 2 divides $|\MAJORITY^{-1}(1)|$.

We first consider the case where $N$ is even, say $N = 2m$. The
$2^{2m}$ possible inputs can be divided into three types:  those with
more 1's than 0's, those with more 0's than 1's, and the ${2m \choose
  m}$ perfectly balanced inputs.  The number of inputs with a majority
of 1's is therefore $2^{2m-1} - \frac{1}{2}{2m \choose m}$.  Since
${2m \choose m} = (2m)!/(m!)^2$, Corollary~\ref{hamming-corollary}
states that ${2m \choose m}$ is exactly divisible by $2^k$ for $k =
(2m - w(2m)) - 2(m - w(m)) = w(2m) = w(N)$. Therefore, since $w(N) <
N$, $|\MAJORITY^{-1}(1)|$ is exactly divisible by $2^{w(N) - 1}$.

If we had $\Dpar(\MAJORITY) \leq N - w(N)$, then, by
Lemma~\ref{rv-lemma}, we would have $2^{w(N)}$ dividing
$|\MAJORITY^{-1}(1)|$.  Since this is false, we must have
$\Dpar(\MAJORITY) \ge N + 1 - w(N)$ for even $N$.

When $N$ is odd, we note that we can use an algorithm for
{\MAJORITY} on $N$ variables to solve the problem on $N-1$
variables:  pad the $N-1$ input bits with one 0.  Since the above
argument for the even case only relies on the number of inputs mapped
to 1, we thus conclude that $\Dpar(\MAJORITY)$ for $N$ odd is at least
$(N - 1) + 1 - w(N - 1) = N + 1 - w(N)$, which proves the desired
result. \qed

\section{Lower Bounds for Computing {\MAJORITY} with Zero-Sided Error}
\label{sec:lower:zero}

Beals et al.~\cite{BBCMdW} prove a lower bound of $\frac{N}{2}$ on
the number of queries a quantum black-box network needs to compute
{\MAJORITY} on $N$-bit strings with zero-sided error $\epsilon < 1$.
We will show that the cost of a randomized {\XOR} decision tree
computing {\MAJORITY} with zero-sided error $\epsilon = o(1)$ 
cannot be reduced below $\frac{2}{3}N - o(N)$. We will also prove
that any classical randomized decision tree with zero-sided error
$\epsilon = \frac{1}{2}$ has to have cost at least $N$. 
In fact, we will show the stronger result that any classical randomized
decision tree with arbitrary error bounded by $\frac{1}{4}$ has 
cost at least $N$.

This result about randomized {\XOR} decision trees follows directly
from the average case lower bound of Alonso et al.~\cite{ARS97}
using a standard argument.

\begin{theorem}[Alonso et al.~\cite{ARS97}]\label{thm-ars-lower-bd}
There exists a constant $d$ such that the following holds for any input 
length $N$.
For any {\XOR} decision tree computing {\MAJORITY}, the
average cost over all inputs of length $N$ is at least 
\[ \frac{2}{3}N - \sqrt{\frac{8N}{9\pi}} - d. \]
\end{theorem}

\begin{corollary}\label{cor:ars:lower}
There exists a constant $d$ such that the following holds for any input 
length $N$.
For any randomized {\XOR} decision tree computing {\MAJORITY} exactly,
there exists an input of length $N$ such that the expected
number of queries is at least $\frac{2}{3}N
- \sqrt{\frac{8N}{9\pi}} - d$ on that input.
\end{corollary}

\proof
Let $g(N)$ denote $\frac{2}{3}N - \sqrt{\frac{8N}{9\pi}} -d$ from
Theorem \ref{thm-ars-lower-bd}.

Look at the randomized {\XOR} decision tree $\T$ as a distribution over
deterministic {\XOR} decision trees $\{\T_i\}$. Each deterministic tree
$\T_i$ in the support of $\T$ computes {\MAJORITY} exactly. By Theorem
\ref{thm-ars-lower-bd}, the average cost of each $\T_i$ is at least $g(N)$.
Consequently, the expected average cost of $\T$ is at least $g(N)$. Therefore,
there exists an input on which the expected number of queries is at
least $g(N)$.
\qed

A randomized {\XOR} decision tree $\T$ with zero-sided error $\epsilon$ and
cost $C$, can be transformed into an exact randomized {\XOR} decision
tree $\T'$ for the same function with an expected number of queries of 
at most $C + \epsilon (N-C) \leq C + \epsilon N$ on any input. 
We just run $\T$ and whenever
it is about to answer ``I don't know,'' we query individual bits until
we know the entire input. Using Corollary \ref{cor:ars:lower}, we obtain:
\begin{theorem}
Any randomized {\XOR} decision tree computing {\MAJORITY} on $N$-bit
inputs with zero-sided error $\epsilon$ has cost at least
$\frac{2}{3}N - \epsilon N - O(\sqrt{N})$.
\end{theorem}

In contrast, a classical randomized decision tree
needs $N$ queries to compute {\MAJORITY} with error $\epsilon$ for
any sufficiently small constant $\epsilon$.
\begin{theorem}\label{thm:lb:two:clas}
Any randomized decision tree that computes the {\MAJORITY} of $N$ bits
with bounded error $\epsilon \leq \frac{1}{4}$ has cost at least $N$.
\end{theorem}
\proof
Let $t$ denote $\ceil{N/2}$. 
Suppose there exists a randomized decision tree $T$ that computes
{\MAJORITY} on $N$-bit inputs with bounded error $\epsilon \leq \frac{1}{4}$ 
and cost at most $N-1$. Without loss of generality, we can assume that 
$T$ always queries exactly $N-1$ of the $N$ input bits.

First the following observations. Consider a deterministic tree of cost
$N-1$ and suppose we pick an $N$-bit input uniformly at random among those 
with exactly $A$ ones. Then the probability that the unique bit not
queried is a one equals $\frac{A}{N}$. Also, for any final state $s$
of $T$, the probability that we end up there only depends on the number
of ones seen when we reach $s$.

Look at $T$ as a probability distribution over deterministic trees of
cost $N-1$. Among all final states that have seen $t-1$ ones, let
$\alpha$ be the weighted fraction that outputs 0. Consider the input
distribution that is a convex combination of $\beta$ times the uniform 
distribution over inputs with exactly $t-1$ ones, and  $1 - \beta$ times the 
uniform distribution over inputs with exactly $t$ ones. By the above 
observations, the probability of error is at least
\[ \beta \left(1 - \frac{t-1}{N}\right) (1-\alpha) + (1-\beta) \frac{t}{N} \alpha.\]
Picking $\beta \doteq \frac{t}{N+1}$ makes the factors of $(1-\alpha)$ and 
$\alpha$ equal, so we get that the probability of error is at least
$\frac{t(N-t+1)}{N(N+1)}$, which exceeds $\frac{1}{4}$. This contradicts
the assumption that $\epsilon \leq \frac{1}{4}$.
\qed

We note that the bound of $\frac{1}{4}$ in Theorem \ref{thm:lb:two:clas}
is essentially tight. Using the notation from the above proof, the following
algorithm does the job: Query $N-1$ bits in random order and output 
0 if less than $t-1$ of them are one, 1 if more, and the outcome of a
(biased) coin toss otherwise.

\begin{corollary}\label{cor:lb:zero:clas}
Any randomized decision tree that computes the {\MAJORITY} of $N$ bits
with zero-sided error $\epsilon \leq \frac{1}{2}$ has cost at least $N$.
\end{corollary}

\proof
Transform the randomized decision tree $\A$ with zero-sided error
$\epsilon$ into the randomized decision tree $\A'$ as follows: Whenever
$\A$ says ``I don't know,'' output the outcome of a fair coin toss;
otherwise answer the same as $\A$. $\A'$ has two-sided error $\epsilon/2$.
Then apply Theorem \ref{thm:lb:two:clas} to $\A'$.
\qed

Again, the bound of $\frac{1}{2}$ is essentially tight.

\section{Open Questions}
\label{sec:questions}
We can summarize the known results in a table. We fixed the error
$\epsilon$ in the table to $N^{-2}$.
\begin{center}
\begin{tabular}{||c||c|c||c|c||} \hline
cost of & \multicolumn{2}{c||}{Quantum black-box model} &
	\multicolumn{2}{c||}{{\XOR} decision tree model} \\ \cline{2-5}
{\MAJORITY} & Lower bound & Upper bound & Lower bound & Upper bound \\ \hline
exact & $N/2$~\cite{BBCMdW} & $N - w(N) + 1$ &
$N - w(N) + 1$~\cite{SW91} & $N - w(N) + 1$~\cite{SW91} \\ \hline
\begin{tabular}{c}zero-sided \\ error \end{tabular}
	& $\frac{1}{2}N$~\cite{BBCMdW} & $\frac{2}{3}N + O(\sqrt{N \log N})$
	& $\frac{2}{3}N - O(\sqrt N)$~\cite{ARS97}
	& $\frac{2}{3}N + O(\sqrt{N \log N})$ \\ \hline
\begin{tabular}{c}two-sided \\ error \end{tabular}
	& $\Omega(N)$~\cite{BBCMdW}
        & $\frac{1}{2}N + O(\sqrt{N \log N})$~\cite{vD98}
	& $\frac{1}{2}N$
	& $\frac{2}{3}N + O(\sqrt{N \log N})$ \\ \hline
\end{tabular}
\end{center}

This leads to several natural open questions.
\begin{itemize}
\item Our results for exact and zero-sided error in the {\XOR} decision
tree model are quite tight. The corresponding results in the quantum
black-box model are not. Can we narrow the gap?  The quantum black-box 
model is more powerful than the {\XOR} decision tree model, so we may
be able to improve the quantum upper bound by applying some other 
technique to {\MAJORITY}.
\item On the other hand, we may be able to improve the quantum lower 
bounds, in particular in the two-sided error case. 
The best lower bound we currently know is $\Omega(N)$ for any
constant error ratio less than $\frac{1}{2}$. This follows from Paturi's
\cite{paturi} result that the approximating degree of the majority 
function (see, for example,
\cite{NS} for a definition) is $\Omega(N)$, and the observation by
Beals et al.\ that half the approximating degree is a lower bound
for the quantum black-box complexity in the bounded error setting.
The constant hidden in the $\Omega(N)$ of Paturi's result is much
smaller than 1. A constant of 1 would show that Van Dam's approach 
is essentially optimal for {\MAJORITY}.
\item In this paper, we focused on the exact and zero-sided error settings.
The results in the table for two-sided error {\XOR} decision trees
trivially follow from the classical lower bound 
(Theorem \ref{thm:lb:two:clas}) and the upper bound in the zero-sided 
error setting (Corollary \ref{cor:GP}). Can we exploit the two-sided
error relaxation?  How about the one-sided error setting?
\item The $O(\sqrt N)$ term in the lower bound for the cost 
of a zero-sided error randomized {\XOR} decision tree comes from
the average size of the discrepancy of a random
input.  It seems likely that, if we restrict to balanced inputs, we
can improve this lower bound to $\frac{2}{3}N - O(\log N)$.  Can we do so?
\end{itemize}

\section{Acknowledgments}
The authors would like to thank L\'aszl\'o Babai, Eric Bach,
Harry Buhrman, Jin-Yi Cai, Ian Dinwoodie, Murali K. Ganapathy, 
Pradyut Shah, Janos Simon, Daniel {\v S}tefankovi{\v c}, Ronald de Wolf,
and the anonymous referees for stimulating discussions, pointers to
the literature, and helpful comments on earlier versions of the paper.

\bibliographystyle{plain}

\countdef\modnum=10 \countdef\yeardiff=11 \newcommand{\ordinal}[1] {{#1}\modnum
  = #1 \divide\modnum by 100 \multiply\modnum by 100 \advance\modnum by 11
  \multiply\modnum by -1 \advance\modnum by #1
  \ifcase\modnum{th}\or{th}\or{th}\else \modnum = #1 \divide\modnum by 10
  \multiply\modnum by 10 \advance\modnum by 1 \multiply\modnum by -1
  \advance\modnum by #1 \ifcase\modnum{st}\or{nd}\or{rd}\else{th}\fi\fi}
  \newcommand{\yearordinal}[2] {\yeardiff=#2 \multiply\yeardiff by -1
  \advance\yeardiff by #1 \ordinal{\number\yeardiff}}
  \newcommand{\STOC}[1]{Proc. of the \yearordinal{#1}{1968} ACM STOC}
  \newcommand{\FOCS}[1]{Proc. of the \yearordinal{#1}{1959} IEEE FOCS}
  \newcommand{\Structures}[1]{Proc. of the \yearordinal{#1}{1985} Annual
  Conference on Structure in Complexity Theory}

\medskip
\par\noindent
{\bf \Large Appendix: Tail Laws }

\medskip

\noindent
In this section, we establish 
Lemmas \ref{lemma:OP} and \ref{lem:Mconcentration} as applications
of Azuma's inequality (see, for example, Motwani and Raghavan~\cite[Section 4.4]{MR97}).

A sequence of random variables $Y_0, Y_1, \ldots, Y_\ell$ is called a
{\em martingale} if $E[Y_i \, | \, Y_0, Y_1, \ldots, Y_{i-1}] = Y_{i-1}$
for every $1 \leq i \leq \ell$. Azuma's inequality is a general tail law
for martingales:

\begin{theorem}[Azuma's Inequality]\label{azuma}
Let $Y_0, Y_1, \ldots, Y_\ell$ be a martingale. 
If $|Y_i - Y_{i-1}| \leq c_i$ for every $1 \leq i \leq \ell$, then 
$$ \Pr \left[ |Y_\ell - Y_0| \geq \lambda \right] \leq 
      2 \cdot \exp\left(- \frac{\lambda^2}{2 \sum_{i=1}^\ell c_i^2}\right)$$
for every $\lambda \geq 0$.
\end{theorem}
If the underlying sample space $\Omega$ can be written as a product
\begin{equation}\label{decomposition}
 \Omega = \prod_{i=1}^\ell \Omega_i, 
\end{equation}
we can associate a random
variable $Y$ with a martingale $Y_0, Y_1, \ldots, Y_\ell$ defined by
\begin{equation}\label{Doob}
Y_i (x_1, x_2, \ldots, x_\ell) = 
    E[\, Y \, | \, X_1 = x_1, X_2 = x_2, \ldots, X_i = x_i]
\end{equation}
for $0 \leq i \leq \ell$, 
where $X = (X_1, X_2, \ldots, X_\ell)$ denotes the sample. The latter
martingale is called the {\em Doob martingale} of $Y$ with respect to
the decomposition (\ref{decomposition}). Note that $Y_0 = E[Y]$ and
$Y_\ell = Y$.

\medskip
\noindent
{\it Proof of Lemma~\ref{lemma:OP}.} \
For simplicity, we assume that $N$ is even; the proof works for odd
$N$ as well.

Consider the Doob martingale $Y_0, Y_1, \ldots, Y_{N/2}$ of $Y = c$
with respect to the decomposition of the sample string in pairs, i.e., 
(\ref{decomposition}) with $\Omega_i = \{0,1\}^2$, $1 \leq i \leq \ell = N/2$. 

Fix $j \in \{1, 2, \dots, N/2\}$, and $x_1, x_2, \ldots x_j \in \{0,1\}^2$. 
The conditional distribution on the right-hand side of (\ref{Doob}) for 
$i=j-1$ can be obtained from the one for $i=j$ by the following transformation:
swap the bit in position $2j-1$ with a bit in a random position 
$p_1$, $2j-1 \leq p_1 \leq N$,
and swap the bit in position $2j$ with a bit in another random position
$p_2$, $2j-1 \leq p_2 \leq N$. Since the transformation affects at most
3 pairs, the value of $c$ can change by no more than 3 units under this
transformation. In fact, a change in $c$ of 3 units is impossible, as it
would require a change in the parity of the bits in the 3 pairs involved,
which is impossible. It follows that $|Y_j - Y_{j-1}| \leq 2$.

Since $E[c] = AB/(N-1)$, Theorem \ref{azuma} yields that
$$\Pr \left[ |c - AB/(N-1)| \geq \lambda \right] \leq
    2 \cdot \exp(-\lambda^2/4N),$$
from which the bound stated in Lemma \ref{lemma:OP} follows.
\qed

\medskip

\noindent
{\it Proof of Lemma~\ref{lem:Mconcentration}.} \
We will first establish a concentration result for the auxiliary
random variables $C_k$, $1 \leq k \leq N$, defined as the number of
ones among the first $k$ positions of the sample string. 

Consider the Doob martingale of $Y = C_k$ with respect to the trivial 
decomposition (\ref{decomposition}) with $\Omega_i = \{0,1\}$, 
$1 \leq i \leq \ell = N$.

Fix $j \in \{1, 2, \dots, N\}$, and $x_1, x_2, \ldots x_j \in \{0,1\}$. 
The conditional distribution on the right-hand side of (\ref{Doob}) for 
$i=j-1$ can be obtained from the one for $i=j$ by swapping the bit in
position $j$ with a bit in a random position in $\{j, j+1, \ldots, N\}$.
The swapping process can affect the value of $C_k$ by at most one for
$j \leq k$, and not at all for $j > k$. It follows that 
$|Y_j - Y_{j-1}| \leq 1$ for $j \leq k$; $Y_j = Y_{j-1}$ for $j > k$.
Note that $E[C_k] = kA/N$. Theorem \ref{azuma} yields that
\begin{equation}\label{bound:C_k}
\Pr \left[ |C_k - kA/N| \geq \lambda \right] \leq
    2 \cdot \exp(-\lambda^2/2k).
\end{equation}

For any $\Delta$, if $\left| M - \frac{N^2}{2A} \right| > \Delta$, then
either
\begin{enumerate}
\item $C_{\frac{N^2}{2A} - \Delta} \geq {N \over 2} = 
  E[C_{\frac{N^2}{2A} - \Delta}] + {A \Delta \over N}$, or
\item $C_{\frac{N^2}{2A} + \Delta} \leq {N \over 2} = 
  E[C_{\frac{N^2}{2A} + \Delta}] - {A \Delta \over N}$.
\end{enumerate}
By (\ref{bound:C_k}), the probability that at least one of the above occurs
is at most
$$
4 \exp \left(-{\left(A \Delta \over N\right)^2 \over 2N}\right) 
= 4 \exp \left (- {A^2 \Delta^2 \over 2 N^3} \right) \leq
4 \exp \left(-{\Delta^2 \over 8 N}\right),
$$
since $A/N \ge 1/2$. The bound stated in Lemma \ref{lem:Mconcentration}
follows.
\qed

\end{document}